\documentclass[aps,pre,twocolumn,floatfix,groupedaddress]{revtex4}
\usepackage{graphicx}  % needed for figures
\usepackage{dcolumn}   % needed for some tables
\usepackage{bm}        % for math
\usepackage{amssymb}   % for math
\usepackage[latin1]{inputenc}
\usepackage{amsfonts}
\usepackage{amsmath}
\usepackage{amsthm}
\usepackage[american,english]{babel}
\usepackage{fontenc}
\usepackage{textcomp}
\usepackage{color}
\begin{document}
%\title{Two-days memory and random global information can help agent coordination: The story of Minority Game with a vivid memory}
%\title{A two-day memory is optimal in minority game with a vivid memory}
%\title{Random global lies can enhance social efficiency: The story of Minority Game with a vivid memory}
\title{Effect of detailed information in Minority Game: Optimality of 2-day memory and enhanced efficiency due to   random exogenous data}
\author{V. Sasidevan}
\email{sasidevan@imsc.res.in,sasidevan@gmail.com}
\affiliation{The Institute of Mathematical Sciences, CIT Campus, Taramani, Chennai
600113, India.}

\date{\today}

%\shorttitle{Minority Game with vivid memory} %Insert here a short version of the title if it exceeds 70 characters

%\author{V. Sasidevan\inst{1}}
%\shortauthor{F. Author \etal}

%\institute{                    
 % \inst{1} The Institute of Mathematical Sciences, CIT Campus, Taramani, Chennai%
%600113, India.\\
 % \inst{2} Second Institute - Address
%}
%\pacs{05.65.+b}{Self-organized systems}
%\pacs{89.75.Fb}{Structures and organization in complex systems}
%\pacs{89.65.Gh}{Econophysics}
%\pacs{05.40.-a}{Third pacs description}

%89.65.Gh, 89.75.Fb, 05.40.-a, 05.65.+b
\begin{abstract}
In the Minority Game (MG), an odd number of heterogeneous and adaptive agents  choose between two alternatives and those who end up on the minority side win. When the information available to the agents to make their choice is the identity of the minority side for the past $m$ days, it is well-known that emergent coordination among the agents is maximum when $m \sim \log_2(N)$. The optimal memory-length thus increases with the system size. In this work, we show that, in MG when the information available to the agents to make their choice is the strength of the minority side for the past $m$ days, the optimal memory length for the agents is always two ($m=2$) for large enough system sizes. The system is inefficient for $m=1$ and converge to random choice behaviour  for $m > 2$ for large $N$.  Surprisingly,  providing the agents with uniformly and randomly sampled $m=1$ exogenous information results in an increase in coordination between them compared to the case of endogenous information  with any value of $m$. This is in stark contrast to the conventional MG, where agent's coordination is invariant or gets worse with respect to such random exogenous information.  
\end{abstract}

\maketitle

\section{Introduction}
\label{sec1}
The Minority Game (MG) is one of the best-known examples of an agent-based model of a complex adaptive system. It was introduced as a variant of  Brian Arthur's El Farol bar problem \cite{arthur_1994, arthur_1999} and from its inception by Challet and Zhang in 1997 \cite{challet_1997}, continues to  attracted significant attention from researchers in diverse areas ranging from physics and computer science to social science and economics \cite{perez_2006, miller_2007,guttmann_2011,helbing_2012, bikas_2014,yang_2014,dong_2014,huang_2015,huang_2015_0,li_2016}.
%One of the best-known examples of such a system is the  \cite{challet_1997,challet_1998}, which was introduced as a variant of  Brian Arthur's El Farol bar problem \cite{arthur_1994}. 
The feature of MG whereby  agents benefit by being in the minority makes it a model of competition for limited resources among agents - an important feature of many systems ranging  from financial markets and traffic to animal foraging \cite{moro_2004}. The model shows a non-trivial collective behavior resulting from a mean-field type interaction between heterogeneous agents. The behavior of the model has been well understood by a variety of analytical frameworks, such as the crowd-anticrowd theory \cite{hart_2001_1,hart_2001_2}, the replica method \cite{challet_2000_1,marsili_2000,marsili_2001} and the generating functional formalism  \cite{coolen_2004,coolen_2006}. 
 
In the classical MG model proposed by Challet and Zhang (henceforth referred to as CZMG),  an odd number of agents $N$ have to select between two alternatives (say A and B) on each round, independently and simultaneously. Those  on the minority side win (each receives  payoff 1) and those on the majority side lose (payoff 0). In CZMG and several of its variants, agents make their selection based on the information about the  identity of the side (whether A or B) that was occupied by the minority group on each of the previous $m$ rounds. A \textit{strategy} is defined as a rule that informs an agent whether to select A or B on the next round for all possible past contingencies. Apart from such strategies  in which agents predict the future minority side based on the identity of the past ones, one may also consider strategies where agents have access to a more detailed information. In this latter class of strategies, agents use the  number of attendees in the minority side in the past  to predict the future 
minority side. Surprisingly, the role of such strategies in MG has never been investigated previously and there exists a perception that using precise information to make a binary prediction is redundant and $N$-dependence of such information is analytically undesirable \cite{challet_2005}. The effect of such strategies on the emergent behaviour of MG is the focus of the present study.

The strategy set in which exact information about the attendance gets mapped to binary decision is also important from a practical point of view. It is conceivable that there are many situations in which an agent not only knows whether she was on the minority side or not, but also how crowded it was relative to the comfort level.   In the language of financial markets, in whose context CZMG is mostly discussed \cite{challet_2005_1}, this means that traders know how many  sellers or buyers were there on  each round. A minority game in which rational and homogeneous agents (as opposed to inductive and heterogeneous agents in CZMG) use such information has been discussed in Ref.~\cite{dhar_2011,sasidevan_2014} (Also see Ref.\cite{reents_2001, biswas_2012,achakraborti}) where  such agents can self-organize to a state with maximum possible global efficiency.

In this work, we show that the class of strategies where the agents use exact information about the past attendance in A  to make a binary decision, which was overlooked earlier because of its  \textquotedblleft undesirable\textquotedblright\; properties, in fact helps the agents to self-organize to an efficient state just as in CZMG, but with notable differences.
We show that when the agents in MG use the exact attendance in A (or B) as the information to predict the future minority side, for large enough system size, their optimal memory-length is 2 independent of $N$. With the information being the attendance in side A, even for  $m=1$, the agents perform well when their number is only a few hundreds. For $m>2$, the efficiency approaches the random choice value as $N$ increases. An intriguing result for the MG with detailed information is that  providing the agents with  uniformly sampled random $m=1$ exogenous information results in a better efficiency than  endogenous information with any value of $m$. Thus, random exogenous information that is commonly provided to all the agents helps them to self-organize better. These results are in stark contrast to the conventional MG with binary information where the optimum memory length increases with system size and the global efficiency remains unchanged  or decreases with random exogenous information.

The rest of the paper is organized as follows. In section \ref{sec2}, we give a precise definition of MG with detailed information. In section \ref{sec3}, we show both analytically and numerically that agent coordination is optimal when they retain the information for two previous rounds. In addition, we  show that random exogenous information results in  enhanced coordination between the agents. We also discuss the effect of different tie-breaking rules that the agents may employ on the global coordination between them.  Section \ref{sec4} contains  discussion and concluding remarks.

\section{MG with detailed information}
\label{sec2}
\begin{figure}
 \centering
   \includegraphics[width = .5\textwidth]{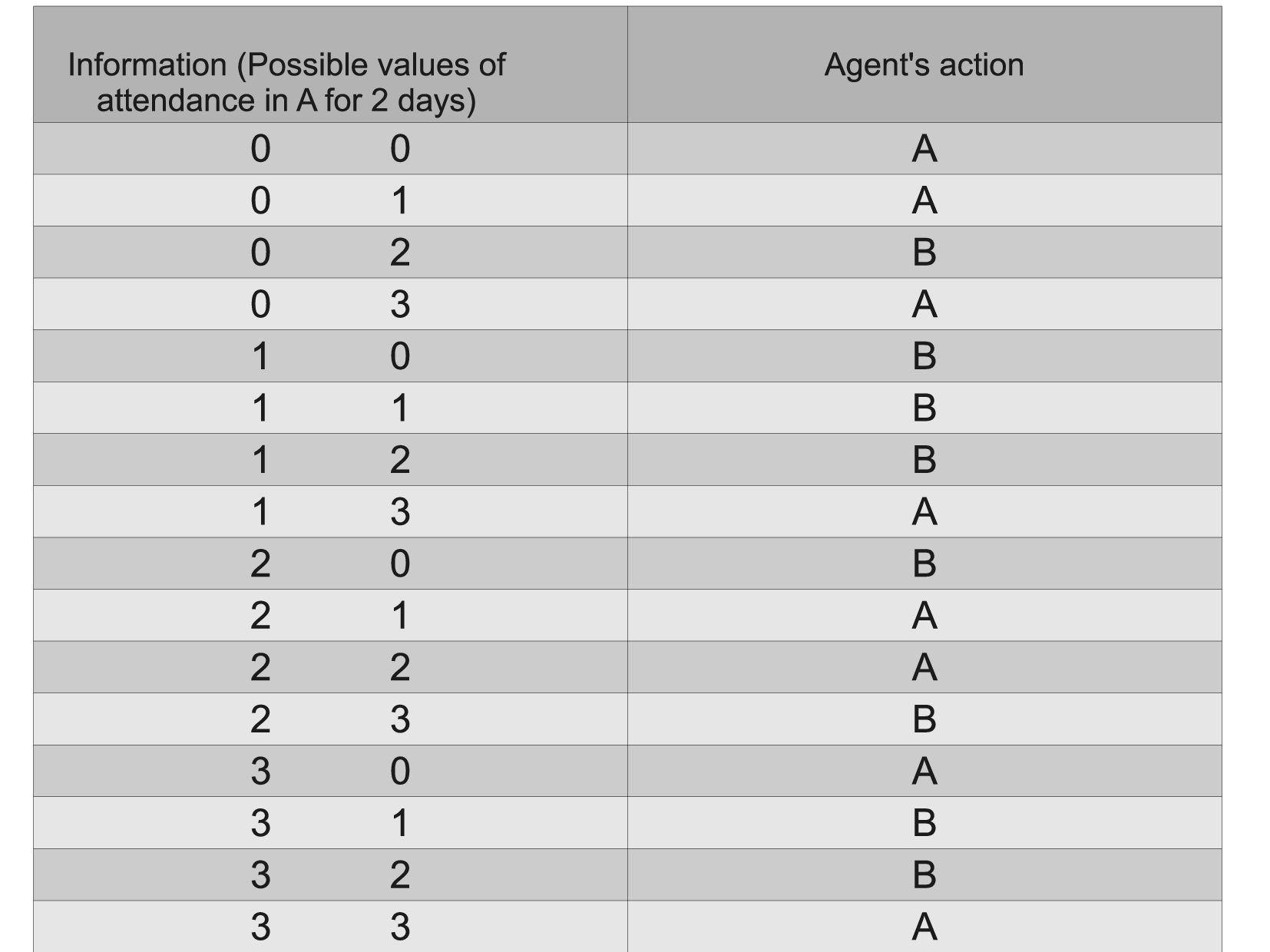} 
   \caption{An example of a strategy for a minority game with $N=3$ and memory-length $m=2$, where agents map precise information in the past (attendance in side A) to a future binary action (choose A or B).} 
 \label{fig_0}
 \end{figure}

The only difference in our present study from CZMG is in the nature of information available to the agents.
There are $N$ agents who select between two alternatives (A and B) on each round, independently and simultaneously. Those  on the minority side receive  payoff 1 and others receive payoff 0. In CZMG, the information available to the agents is the identity of the minority side on each of the previous $m$ rounds. A \textit{strategy} is a rule that maps the information (a binary string of length $m$) to future action (choose A or B). The total number of strategies is therefore $2^{2^M}$. In our present problem, information available to the agents is the strength of the minority side  on each of the previous $m$ rounds. A \textit{strategy} is a rule that maps the exact attendance in the past $m$ rounds to future binary action (An example of such a strategy for $N=3$ and $m=2$ is shown in Fig.~\ref{fig_0}). It is easy to see that the total number of strategies in such a case is  $2^{(N+1)^m}$.

The time evolution of the game is exactly as in CZMG: At the beginning, each agent is given a small number  of strategies (say 2) randomly selected from the full set of possible strategies. During the game, each agent measures the performance of her assigned strategies by keeping track of how well they predicted the minority in the past (by keeping a score for each of her strategies). On a given round, each agent uses her best performing strategy, i.e. the strategy that has correctly predicted the minority side the most number of times until that round. We are interested in the collective behaviour of the system when agents use such strategies.

\begin{figure}
\centering
  \includegraphics[width = .5\textwidth]{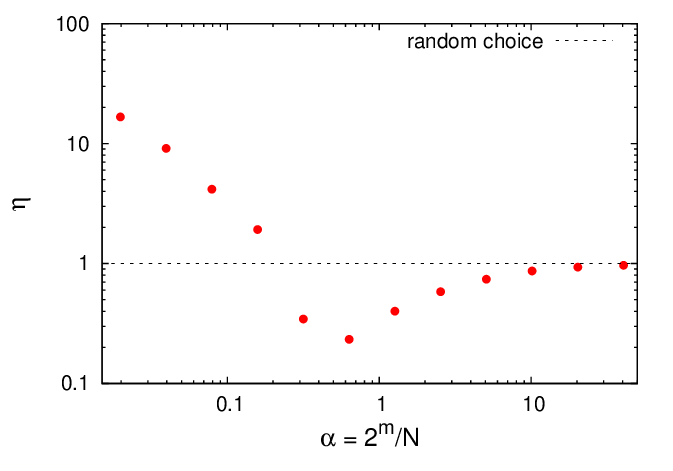} 
  \caption{Variation of the inefficiency $\eta$ (Eq.~\ref{eq1}) with the parameter $\alpha = 2^m/N$ for CZMG with two strategies per agent. The line $\eta = 1$  corresponds to the case where agents select randomly between the two alternatives A and B. Each data point is obtained by averaging over 100 realizations.} 
\label{fig_1}
\end{figure}

\section{Results}
\label{sec3}
In a minority game, the global inefficiency  which measures the strength of fluctuations in the attendance in side A (or B) about its average value $N/2$ can be quantified by the inefficiency parameter $\eta$,
\begin{equation}
 \eta = \frac{4}{N} \langle (r -N/2)^2 \rangle,
 \label{eq1}
\end{equation}
where $r$ is the attendance in A and $\langle ~\rangle$ denotes averaging over a long time evolution in the steady state, and over different initial conditions. The normalization has been chosen so that the inefficiency parameter $\eta$ of the system with agents  selecting  randomly between A and B is $1$. 
% A significant feature of CZMG was that heterogeneous and competing agents
% gave rise to non-trivial collective phenomena, in that they could self-organise into a globally efficient state where resources are better utilized compared to the case in which agents selected randomly between A and B. 

The most interesting feature of CZMG is that for large enough values of the memory length $m$, agents self-organize  into a state where the fluctuation in the attendance of a side about its mean value ($N/2$ because of the symmetry between A and B)  is minimized.  Thus in CZMG,  selfish agents who care only about their personal gain self-organize into a globally efficient state where the utilization  of resources is maximized in comparison to a simple random choice behavior by the agents where they select A and B with equal probability.
In CZMG, it was found that for large enough memory length $m$, $\eta \ll 1$  for a given value of $N$ (See Fig. \ref{fig_1}). The value of $m$ at which $\eta$ is a minimum is a function of $s$ - the number of strategies per agent - but is approximately given by $2^m \approx N$. Thus the critical memory length, say $m_c$, is an increasing function of the number of agents and is given by $m_c \sim \log_2 N$ \cite{challet_1998} (the precise value of $m_c$ can be obtained analytically, see for e.g. Ref.~\cite{challet_2000_1,challet_2005}). The optimal memory-length thus increases with the system size. For memory lengths that are much smaller than $m_c$, it was found that the system shows herding behavior where a large fraction of agents choose the same option simultaneously. In this case,  the global efficiency and individual payoffs of agents are extremely low and become much worse than simple random choice behavior by the agents. 

%is highly inefficient, characterized by very high values of $\eta$ where the average payoff of agents is extremely low. For $m \ll m_c$  the system shows herding behavior where a large fraction of agents choose the same option simultaneously. In this case,  the global efficiency and individual payoffs of agents are extremely low and become much worse than simple random choice behavior by the agents. 

Now we will study the behavior of the inefficiency parameter $\eta$ when agents have access to the number of attendees on the minority side for the past $m$ days. As already mentioned in the introduction, it was argued in Ref.~\cite{challet_2005} that if agents use precise information about the attendance in A to make a binary decision as in our model, it has  issues of redundancy and an undesirable dependence on system size $N$. We will show that both these issues are irrelevant for agent coordination and in fact the precise information helps the agents to achieve self-organization into a globally efficient state for a fixed memory length 2, regardless of the size of the system $N$. 

One of the early results that gave much insight into the behavior of CZMG was the existence of a reduced strategy space (RSS) in which any two strategies  are either  uncorrelated or anti-correlated \cite{challet_1998}. It was shown that the behavior of CZMG is unchanged if, instead of the full strategy set which is $2^{2^m}$ in number, one use a reduced strategy set that contains $2\times 2^{m}$ strategies.  When the  number of strategies in the RSS is much higher than the number of agents, i.e, when $2\times2^m \gg N$, we get essentially a random-choice behavior with $\eta \rightarrow 1$ ($\alpha \gg 1$ in Fig.~\ref{fig_1}) and when $2\times2^m \ll N$, a significant fraction of agents share the same set of strategies and we get very high fluctuations in the attendance ($\alpha \ll 1$ in Fig.~\ref{fig_1}). Importantly, for $2\times2^m \approx N$ (or when $\alpha \sim {\cal O} (1)$ in Fig.~\ref{fig_1}), the fluctuation is minimized and we get a highly efficient system (for $s=2$, the inefficiency $\eta$ 
is minimum at $\alpha \approx  0.3374$ \cite{challet_2000_1}).

\begin{figure}
\centering
  \includegraphics[width = .5\textwidth]{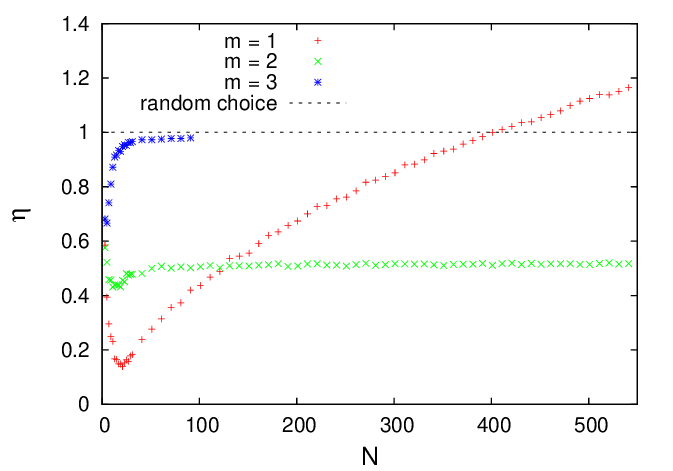} 
  \caption{Variation of inefficiency $\eta$ with system size $N$ for different values of the memory-length $m$, when agents map exact information to binary action in the minority game.  $\eta = 1$  corresponds to the case where agents select randomly between the two alternatives A and B. Each data point is obtained by averaging over 1000 time steps in the steady state and  250 game realizations.} 
\label{fig_2}
\end{figure}

Now, with the information being the exact attendance in A, the  full strategy space has $2^{(N+1)^m}$ elements. This is exponentially increasing with $N$ for all values of $m$,  and therefore  we expect the fluctuations in attendance to  be always  of the order of $\sqrt{N}$ for large $N$ (the following arguments are hence valid in the large $N$ regime and will be supplemented by simulation results for the small $N$ behavior). This has the consequence that  many entries in a strategy table are irrelevant.  Of the $(N+1)$ possible values of the attendance between $0$ and $N$, only $\sqrt{N}$ on either side of the average attendance $N/2$ are relevant. So the full set of strategies only contains about $2^{\left(\sqrt{N}\right)^m}$ relevant entries. Now, just as in CZMG,  we can construct an RSS out of this relevant set, in which any two strategies are either uncorrelated or anti-correlated  with each other (strictly speaking, the arguments to construct the RSS 
in Ref.~\cite{challet_1998} will go through in our model only when $N^{m/2}$ is a power of 2. However, here we are interested only in the scaling behavior of the number of strategies in the RSS with $N$).  It is easily seen that the number of strategies in the RSS  in our model is $P \sim  N^{m/2}$. The behavior of the model thus depends on the ratio $P/N$. Three behaviors are possible: i) 
For $m = 1$, $P/N \sim 1/\sqrt{N}$. The number of strategies  becomes much smaller than the number of agents for large $N$, and hence the inefficiency of the system keeps increasing with $N$. For small $N$, it may still happen that the system has $\eta < 1$ which is indeed what we find from simulation results (see Fig.~\ref{fig_2}). ii) For $m=2$, $P/N \sim {\cal O}(1)$, independent of $N$. The  behavior is exactly the same as  the most efficient phase of CZMG  and hence we expect the system to have a constant $\eta < 1$, independent of the value of $N$. iii) For $m >2$, $P/N$ is strictly increasing with $N$. Thus, the number of strategies in RSS outweighs the number of agents and  we expect that the system will have the random choice behavior characterized by $\eta \rightarrow 1$ as $N$ is increased.

The results for the inefficiency parameter $\eta$ as a function of the system size $N$ from Monte Carlo simulations is shown in Fig.~\ref{fig_2}. For simplicity, we consider the case where there are two strategies per agent. We can see that for $m=2$, the system  always has $\eta <1$ and for large values of $N$,  $\eta \approx 0.5$. For $m=1$, the system has $\eta < 1$ for relatively small number of agents, but becomes inefficient for $N \gtrsim 400$ or so. For $m=3$ and above (only $m=3$ is shown), the system has $\eta \rightarrow 1$ as predicted.

\begin{figure}
\centering
  \includegraphics[width = .5\textwidth]{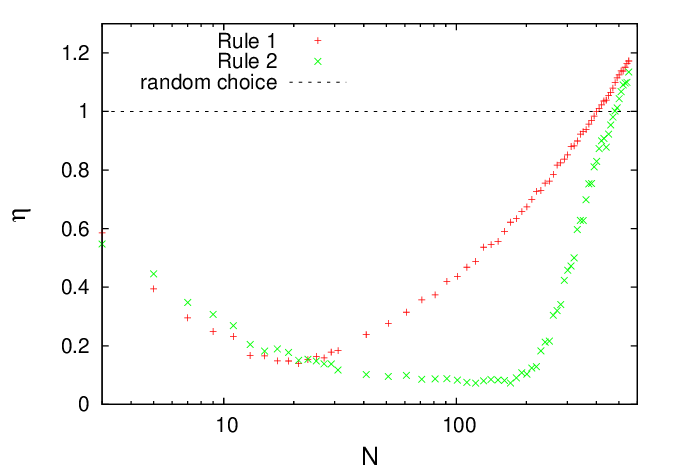} 
  \caption{Variation of inefficiency $\eta$ with system size $N$ for two different tie-breaking rules (see text) and memory-length $m=1$.} 
\label{fig_3}
\end{figure}

An important property of CZMG is that  most of its qualitative features  are independent of the source of the information \cite{cavagna_1999,challet_2000}, i.e. even if the agents are given a uniformly and randomly sampled $m$ bits as information in each round, the behavior of the model remains the same.  In other words,  all the agents in CZMG react to the same piece of information, and it does not matter whether this information is endogenous or a random exogenous one. For CZMG, with uniformly and randomly sampled exogenous information, while  the global inefficiency $\eta$  remains invariant for $\alpha \leq \alpha_c$, it increases a bit for $\alpha > \alpha_c$ \cite{challet_2000}.  In our problem, with the information being the exact attendance in side A, it is easy to deduce that when the agents are given a uniformly and randomly sampled integer between $0$ and $N$ as the information in each round, the problem maps exactly to CZMG (with uniformly and randomly sampled exogenous 
information) 
with $m \sim \log_2 N$. This is because we need $\log_2 N$  bits to represent an integer of order $N$. This implies that the agents will self-organize to a highly efficient state even when $m=1$ if they are provided with such artificial random exogenous information instead of information which is endogenous. Again, this behavior is independent of the system size $N$. Indeed, it is found from simulation studies that, with $m=1$ uniformly random exogenous information, we get $\eta \approx 0.36$. Note that, with $m=2$ endogenous information, for large $N$,  $\eta \approx 0.50$ (Fig. \ref{fig_2}), which is higher than that with  $m=1$ random exogenous information. This can be intuitively understood based on the fact that, providing the agents with uniformly random exogenous information essentially increases the number of strategies in RSS from $\cal{O}(\sqrt{N})$ to $\cal{O}(N)$ [here the Gaussian distribution for the attendance  in side A is replaced by a uniform distribution. One may also consider  exogenous 
information with other possible distributions.
If the distribution \textquoteleft stretches\textquoteright\; the original Gaussian, it will lead to better global efficiencies as it will essentially increase the number of strategies in RSS. As the uniform distribution is the one which does the maximum \textquoteleft stretching\textquoteright\;, it should result in maximum global efficiency among the possible distributions]. Also, with a uniformly random exogenous information of length $m$, it is easily seen that the RSS contains $N^m$ strategies, thus making $m=1$ the optimal choice.  Thus, for large $N$, with exact attendance as information in MG, a uniform, random $m=1$ exogenous information results in a better efficiency than endogenous information with any value of $m$.

\begin{figure}
\centering
  \includegraphics[width = .5\textwidth]{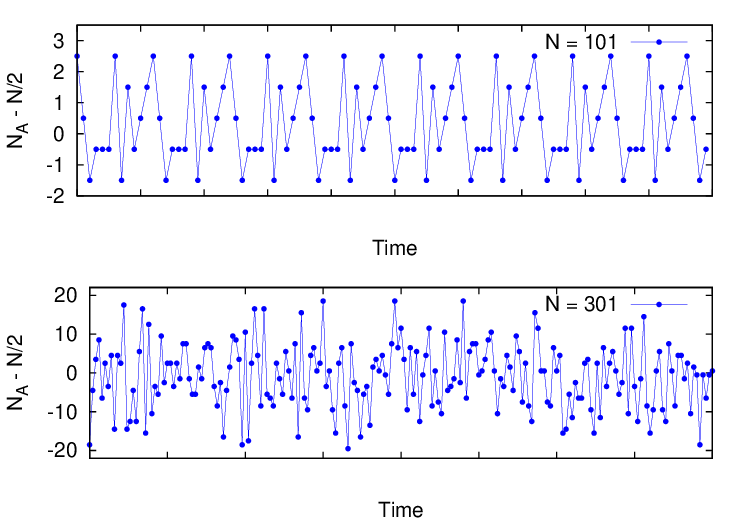} 
    \caption{Typical time series with Rule~2 (see text) for system sizes $N=101$ and $N=301$. Here, $N_A$  denotes the attendance in side A. Periodic fluctuations are seen for a significant range of values of $N$, till about $N\approx 250$.  Successive data points are joined by a line to guide the eye.} 
\label{fig_4}
\end{figure}

Another feature of  CZMG is that the inefficiency curve in Fig.~\ref{fig_1} is not affected by the precise nature of the rule that the agents use to break ties between strategies that have equal scores [Note that in most of the later studies in MG, agents selects stochastically between her strategy-tables. See Sec. \ref{sec4} for a discussion of this. The question of tie-breaking  does not arise in such a case]. For example, let us consider two plausible tie-breaking rules,

\noindent
{\it Rule~1}: When two strategy-tables have equal score, an agent will randomly select one of them.\\
{\it Rule~2}: When two strategy-tables have equal score, an agent will stick to the previously used one.~Thus the agents will change their current strategy-table only when another strategy-table gets a higher score.
\noindent

{\it Rule~1} is what we have used so far to obtain our results.  With exact attendance as information, though the qualitative nature  of the inefficiency parameter remains the same with both these rules, there are significant differences. The variation of $\eta$ with these two rules for $m=1$ is plotted in Fig.~\ref{fig_3}. As we can see, {\it Rule~2} leads to better overall efficiency compared to {\it Rule~1} for a significant range of values of $N$. It is interesting to compare typical time series with {\it Rule~2} for low and high values of $N$ (shown in Fig.~\ref{fig_4}). As we can see, we get periodic oscillations in the attendance for the smaller  $N$ (these oscillations seem to persist for a significant range of values of $N$, till about $N\approx 250$). This is remarkable, since in CZMG a periodic variation in the attendance is  always associated with very low efficiencies (periodic fluctuations are seen in CZMG for $\alpha \ll 1$ regime in Fig.~\ref{fig_1}) whereas here it is associated with very 
high efficiency. 

\section{Discussion and concluding remarks}
\label{sec4}
In this work, we have studied the emergent behavior in Minority Game when the agents use the information about the number of attendees in the minority side in the past  to predict the identity of the future minority side. Such strategies lie at the interface of those considered in CZMG and the El Farol bar problem which inspired it (In the El Farol bar problem, agents use information about the number of attendees in the minority side in the past to predict the the strength of the future minority side. The prediction of exact bar attendance is redundant as an agent's optimum decision is always to choose the minority side regardless of the number of agents there). Using scaling arguments and by numerical simulations, we showed that the agents  who use the exact size of the minority in the past to predict the future minority side will give rise to high global efficiency, except for the case when a large number of agents  retain the information only from the previous round.  In particular, agents with two-days 
memory will self-organize to have a better than random choice efficiency for all system sizes making it the optimal memory length for efficient resource 
sharing. If the agents are provided with uniformly and randomly sampled  exogenous information, the global efficiency becomes better. 
Thus random global information enhances efficiency of resource sharing  when agents have a detailed memory. The theoretical arguments presented depend only on the scaling of the size of the Reduced Strategy Space with the number of agents and hence the conclusions should hold for simple variations of the problem one can think of (for e.g. considering more number of strategies per agent).

Though the Minority Game is introduced and studied as a model where inductively rational agents with limited information could self-organize to an efficient global state, it is worthwhile to compare the performance of such agents with agents who are deductively rational. 
%The optimal strategy of a deductively rational agent, i.e. the Nash equilibrium, depends upon the assumptions made about the game such as the type of information about the past outcomes  available to the agents (or whether there is any information available at all). 
For the Minority Game in which information about the past outcomes is not  available to the agents, the optimal strategy of a deductively rational agent, i.e. the Nash equilibria, were derived in Ref. \cite{marsili_2000,marsili_2001_1}. Apart from the symmetric Nash equilibrium, where agents choose A and B with probability 1/2, there exists asymmetric Nash equilibria  in which $2k$ agents play pure strategies with $k$ of them choosing A, $k$ of them choosing B and the rest $N-2k$ choosing randomly between A and B with equal probability. Note that, of these equilibria, the one with $k = (N-1)/2$ corresponds to the most globally efficient state. For the conventional MG where agents make use of the information about the identity of the past minority sides  to choose between the two sides,  the  problem of finding the Nash equilibria has been addressed in Ref.  \cite{marsili_2000,marsili_2001_1,demartino_2001}.  
%When it comes to agents who use information from the past to decide between A and B, 
An important consideration in studying whether the inductively rational agents in the conventional MG can converge to Nash equilibria is whether they account for market impact or not, i.e. whether each agent accounts for her own contribution to the aggregate outcome in each round of the game \cite{marsili_2000}. This problem is generally addressed in the case where information is assumed to be random and exogenous and %the fact that the choice she makes has an effect on the outcome of the game 
%for the fact that the strategy they employ has a less chance of predicting the minority correctly than the strategies they don't employ. 
it is known that  while the agents in CZMG who do not account for market impact will not be able to converge to Nash equilibria,  a game in which the agents do account for their market impact will converge to Nash equilibria  with very high global efficiencies \cite{marsili_2000,marsili_2001_1,demartino_2001,marsili_2001_a}.  

The problem of deductively rational agents in Minority Game who use the information about the exact size of the minority in the past  to choose between the two sides has been discussed in detail in Ref.~\cite{dhar_2011,sasidevan_2014}. In this case, it was shown that the Nash equilibria  constitute a win-stay lose-shift type strategy, using which the agents were  able to converge to the maximal efficient state ($k = (N-1)/2$ as described above) very quickly. The resulting outcome however is a \textquoteleft trapping state\textquoteright\; in which majority of the agents will continue to remain on the losing side forever. The problem was resolved using a Co-action framework  for deductively rational agents \cite{sasidevan_2014,sasidevan_2015}. It is clear from the results described in the previous section that while the inductively rational agents in our present study who use strategy-tables such as that in Fig.\ref{fig_0} to make their future choice were able to perform better than in the symmetric Nash 
equilibrium, they were not able to self-organize to the best possible equilibria. Whether accounting for market impact will result in an increase in global efficiency is an open problem. As we already noted, for the conventional MG, the question of market impact is usually addressed in a setting where information is assumed to be uniformly and randomly sampled exogenous histories and in such a system, agents accounting for market impact can increase the global efficiency \cite{marsili_2000,marsili_2001_1,demartino_2001}. As shown in Sec. \ref{sec3}, our problem with uniformly random exogenous information maps exactly to CZMG with uniformly random exogenous information  and hence the former result should carry over even with the inclusion of market impact. However with endogenous information, it is not clear  whether accounting for market impact will result in an increase in global efficiency and whether the steady state will correspond to Nash equilibria. This issue is worth investigating, considering the 
fact that, though the agents in our problem with $m=2$ self-organize to a better than random choice behavior, the global efficiency thus achieved is much less than the maximum efficiency achieved in CZMG (compare the minimum in Fig. \ref{fig_1} with the $m=2$ curve in Fig. \ref{fig_2}). We only note that, in this case, we will have to deal with the additional complication that various possible histories (information about the strength of minority in the past) in our problem do not have a uniform distribution.

In this work, we have assumed that in each round of the game, agents use their best-performing strategy, i.e, the strategy that has accumulated the most number of points up to that round, to make their future choice. We may generalize this and consider agents who select probabilistically between their strategies in each round of the game. In most studies of the conventional MG, it is generally assumed that the agents select the strategy to use in a given round of the game probabilistically according to a {\it logit} learning scheme \cite{cavagna_1999_1}. In this, in each round of the game,  an agent $i$ chooses between the strategy-tables $s$ in her hand  in a probabilistic fashion according to,
\begin{equation}
 Prob\{s_i(t) = s\} \propto \exp ^{\Gamma U_{i,s}(t)},
\end{equation}
where $U_{i,s}(t)$ is the accumulated score of strategy  $s$ of agent $i$ at time $t$ and $\Gamma$ is an \textquoteleft inverse temperature\textquoteright\; which relates to the speed of learning \cite{marsili_2001_1}. In the conventional MG, it is known that the parameter $\Gamma $ in the learning scheme has an effect on the global efficiency only at low value of the parameter $\alpha$ where a low $\Gamma$ decreases fluctuations in the attendance and thereby increases global efficiency \cite{cavagna_1999_1,hart, challet_2000_2} (Fig. \ref{fig_1} shows the variation of inefficiency of conventional MG with $\alpha$ for $\Gamma \rightarrow \infty$). The case we discussed in the present work has $\Gamma \rightarrow \infty$, where an agent always uses her best performing strategy. One would  anticipate that, in our problem only the $m=1$ (memory-one) result is going to be influenced significantly by the introduction of a learning rate $\Gamma$. This is because, only the  $m=1$ case corresponds to the regime $\alpha < 1$ of CZMG  where the number of agents outweighs the number of strategies in the Reduced Strategy Space (see Sec. \ref{sec3} for details). The case $m=2$ corresponds to the critical point $\alpha \sim 1$ in CZMG and $m >2$ corresponds to $\alpha > 1$ where the global efficiency does not depend upon $\Gamma$.

More generally our study invokes questions regarding the influence of different kinds of information, and predictions thereof, on the collective statistical properties of complex adaptive systems like the MG.  The CZMG in which agents use binary histories and the present case in which the agents use the  exact attendance histories, represent two extreme cases of information, i.e. fully coarse-grained and fully detailed, respectively. As we saw, the choice of the type of information or the level of  coarse-graining of past data, can have a significant effect on the collective properties of the system. One may also consider an intermediate level of coarse-graining of information between the two extreme cases described above and it will be instructive to study its  effect on the emergent properties.  More generally, one can imagine a system in which agents who use different levels of coarse-graining of the past data (histories) coexist. The simplest setting would be to consider a scenario in which a 
subpopulation of agents use the identity of the past minority sides to predict the identity of the future minority side  and the other subpopulation use the attendance of the past minority sides for the same purpose.

It will be worthwhile to look into the rich literature on Minority Game and its variants in light of the  strategies discussed here, where agents map exact information in the past to future binary action.

\acknowledgments
I would like to thank Sitabhra Sinha for helpful discussions. I would also like to thank Renjan John, Anvy Moly Tom and Shakti N Menon for useful comments on the manuscript.  I thank the HPC facility at IMSc for providing access to \textquotedblleft Satpura\textquotedblright\;.

\end{document}